\documentclass[aps,10pt,twocolumn,tightenlines]{revtex4}
\usepackage{graphics}
\usepackage{graphicx}
\usepackage{pstricks}
\newcommand{\be}{\begin{eqnarray}}
\newcommand{\beq}{\begin{equation}}
\newcommand{\eeq}{\end{equation}}
\newcommand{\ee}{\end{eqnarray}}

\newcommand{\bmp}{\noindent\begin{minipage}{16cm}}
\newcommand{\emp}{\end{minipage}\vskip 7mm} 

\newcommand{\lsim} {\buildrel < \over {_\sim}}
\newcommand{\gsim} {\buildrel > \over {_\sim}}
\newcommand{\al}{\alpha}
\newcommand{\N}{{\cal N}}
\usepackage{bm}
\usepackage{amsmath}
\usepackage{amsfonts}
\usepackage{bbm}

\begin{document}
\title{Color dipole cross section and inelastic structure function}

\author{Yu Seon Jeong\footnote{yusjeong@yonsei.ac.kr} and C. S. Kim\footnote{cskim@yonsei.ac.kr}}
\affiliation{Department of Physics and IPAP, Yonsei University, Seoul 120-749, Korea}
\author{ Minh Vu Luu\footnote{minh-luu@uiowa.edu} and Mary Hall Reno\footnote{mary-hall-reno@uiowa.edu}}
\affiliation{Department of Physics and Astronomy, University of Iowa, Iowa City, Iowa 52242}

\date{\today}

\begin{abstract}
\noindent Instead of starting from a theoretically motivated form of the color dipole cross section in the dipole picture of deep inelastic scattering, 
we start with a parametrization of the deep inelastic structure function for electromagnetic scattering with protons, and then extract the color dipole cross section. 
Using the parametrizations of $F_2(\xi=x\ {\rm or}\ W^2,Q^2)$ by Donnachie-Landshoff and Block et al., 
we find the dipole cross section from an approximate form of the presumed dipole cross section convoluted with the perturbative photon wave function for virtual
photon splitting into a color dipole with massless quarks. The color dipole cross section determined this way 
reproduces the original structure function within about 10\% for $0.1$ GeV$^2\leq Q^2\leq 10$ GeV$^2$. We discuss the large and small form of the dipole cross section and compare with other parametrization.

\end{abstract}

\pacs{13.60.Hb, 12.38.Bx}
\maketitle

\section{Introduction}

Of broad interest in particle and astroparticle physics is the hadronic cross section.
The inclusive hadronic cross section is measured over a range of energies
in laboratory experiments \cite{pdg}, and in cosmic ray experiments,
for example, the recent proton-air cross section measured at $\sqrt{s}=57$ TeV
by the Pierre Auger Observatory \cite{pao}.
The hadronic cross section is important for the analysis of a range of experiments
and it is an essential input to many particle shower simulations,
particularly for the forward production of hadrons.
In the context of astroparticle physics, the hadronic cross section is needed to
calculate the atmospheric flux of leptons from cosmic ray interactions
with air nuclei \cite{gaisser,lipari}.

The structure of hadrons is also probed by weak and electromagnetic interactions.
Laboratory experiments in electron scattering and neutrino scattering with hadronic targets
measure cross sections up to $\sqrt{s}\simeq 200$ GeV \cite{pdg},
while neutrino telescopes and gamma ray telescopes are sensitive to even higher energies.
For purely hadronic interactions and electroweak inelastic scattering with hadrons,
in many cases, one is interested in kinematic regimes far different from the experimentally measured regimes.
This requires additional theoretical input to the modeling of the hadronic structure.

Hadronic structure is most easily studied with deep inelastic electromagnetic scattering.
Deep inelastic electromagnetic scattering is characterized by structure functions $F_i$
that depend on kinematic variables Bjorken-$x$ and momentum transfer $Q$.
In the parton model, the structure functions are written in terms of parton distribution functions
with an evolution in $\ln Q^2$ governed by the Dokshitzer-Gribov-Lipatov-Altarelli-Parisi (DGLAP) equations
\cite{gribov, altarelli, dokshitzer}.
When the Bjorken-$x$ range becomes very small, DGLAP evolution fails.
In parton language, the failure is due to gluon recombination
which introduces nonlinear effects into the evolution equations \cite{glr,mq}.
More generally, unitarity is ultimately violated with the DGLAP approach.

An alternative approach which models small $x$ saturation effects is the color dipole model
\cite{nikolaev, mueller}.
The dipole model for DIS has virtual
photons fluctuate to $q\bar{q}$ dipoles which then scatter with the hadronic target via the dipole cross section $\hat{\sigma}$. In this
approach, the structure function $F_i$ comes from the convolution of the virtual photon wave function squared (for splitting into the color dipole) $|\Psi_{L,T}^{(f)}(r,\alpha ;Q^2) |^2$ with the dipole cross section $\hat{\sigma}(r,\xi)$, integrated over the spatial splitting of the dipole $r$ and the longitudinal momentum fraction $\alpha$ of the quark in the dipole.
The quantity $\xi$ may be Bjorken-$x$ or the square of the total energy $W^2$ in $\gamma^*p$ collisions.
The dipole picture can be useful in many contexts, for example, ultrahigh energy neutrino scattering where one may approach the unitarity limit, and high energy hadronic production of charm, where small $x$ and $Q\sim m_c$ are relevant kinematic variables.

The usual starting point in the color dipole model is to postulate a functional form of the dipole cross section guided by
theoretical or phenomenological considerations.
Given this parametrization of $\hat{\sigma}$, hadronic processes as well as DIS scattering can be used to determine free parameters in
$\hat{\sigma}$.
There are a number of models for the dipole cross section \cite{gbw,sgbk,bgbk,iim,soyez,albacete,aamqs,forshaw,kowalski,gkmn}.
The Golec-Biernat-W\"ustoff model \cite{gbw} explicitly includes saturation and
an effect called geometric scaling. Geometric scaling describes the behavior of DIS observables
at low $x$, for example $\sigma_{\gamma*p}$, depends
on a scaling variable ${\cal T}=Q^2 R_0^2(x)$, independent of $W^2$ \cite{sgbk}.
Other authors, including Soyez \cite{soyez} have done more elaborate fits to DIS data
for dipole cross section parametrization guided by theory that also include geometric scaling
\cite{bgbk,iim}.

Beside the parametrization of the dipole cross section, there are direct parametrization of $F_2(x,Q^2)$
\cite{dl,haidt0,block1,block2,block3,block4,block5,haidt,block,blocknew}
based on fits to the DIS data \cite{h197,zeus99,zeus00,h101,zeus01,zeus},
typically for small $x$ and moderate $Q^2$.
The functional form of the $F_2$ fits are postulated independent of any guidance from dipole cross sections.

Our goal here is to start with a parametrization of $F_2$, then approximately invert it to find a corresponding $\hat{\sigma}$. This is a purely phenomenological approach: as discussed below, we are only able to 
make an inversion to find  an approximate $\hat{\sigma}$.
Eventually, one hopes to explore how
different treatments of the small $x$ behavior of $F_2$ guided by unitarity considerations translate to the dipole cross section and vice versa.  This paper is a first step in this phenomenological program.

We take advantage of approximate relations between $F_2$ and $\hat{\sigma}$ as discussed by
Ewerz, von Manteuffel and Nachtmann in Ref. \cite{ewerz}.
As noted by Ewerz et al. \cite{ewerz}, the convolution of the photon wave function and the dipole cross section is simplified in the $m_f=0$ case. We use
Fourier transforms to factorize the convolution for $m_f=0$ to find an approximate $\hat{\sigma}_0$ from an
inverse Fourier transform. 
We do not invert the convolution for massive quarks. However, with a multiplicative normalization factor, we find that 
 $\hat{\sigma}_0$ nearly reproduces $F_2$ even  when $m_f\neq 0$. This allows us
to compare our numerical results with the small $Q$ and large $Q$ approximations of Ref. \cite{ewerz} to gain more insight into the
interplay between the dipole cross section and the electromagnetic structure function. 

In the next section, we review the relevant formulas for the structure function $F_2$ in terms of the dipole cross section and photon wave functions. We show the convenient variables introduced in Ref. \cite{ewerz} make the problem amenable to inversion by Fourier transforms in the case of $m_f=0$. In Sec. III, we discuss the determination of $\hat{\sigma}_0$ in toy model with a single massless unit charged quark.
Fourier transforms require knowing the structure function at both low and high $Q^2$. We use the Donnachie-Landshoff (D-L) parametrization \cite{dl} of $F_2$ to illustrate the procedure in the toy model.  
The D-L parametrization is defined for the full range of $Q^2$, even if it does not describe DIS data for the full $Q^2$ range.
In Sec. IV, we look at the more physical case with 5 massive quarks for both the D-L \cite{dl} parametrization and the new parametrization of 
$F_2$ by Block et al. in Ref. \cite{blocknew}. We find that with a simple normalization constant independent of $x$, the dipole cross section $\hat{\sigma}_0$ does quite well in reproducing the Donnachie-Landshoff structure function even in the massive case. For the Block et al.
parametrization, the normalization depends on $x$ or $W^2$ and reproduces the original $F_2$ parametrization to within about 10\% for
$Q^2=0.1-10$ GeV$^2$. We compare our results with the GBW \cite{gbw} and Soyez \cite{soyez} dipole cross section parametrization and with the small $Q$ and large $Q$ approximations of Ref. \cite{ewerz}.
We conclude in Sec. V.

\section{Dipole Formalism and the Dipole Cross Section}

The dipole formula for deep inelastic electromagnetic scattering involves the photon wave function
and the dipole cross section. Our focus is on the electromagnetic structure function $F_2(x,Q^2)$ in deep inelastic scattering where $ep\to eX$ involves the subprocess
$\gamma^* p\to X$. In terms of the subprocess cross sections, the dipole picture has \cite{nikolaev}
\begin{equation}
F_2(x,Q^2)=\frac{Q^2}{4\pi^2 \alpha_e}\Biggl[ \sigma_L^{\gamma * p}(\xi,Q^2) + \sigma_T^{\gamma * p}(\xi,Q^2)\Biggr]
\end{equation}
in terms of the virtual photon-proton cross section
\begin{eqnarray}
\nonumber
\sigma_{L,T}^{\gamma * p}(\xi,Q^2) &=&\sum_f\int d^2r\int_0^1 d\al \mid \Psi_{L,T}^{(f)}(r,\al;Q^2) \mid^2\\
&\times& \hat{\sigma}(r,\xi)\ .
\label{eq:conv}
\end{eqnarray}
The photon wave functions are written in terms of the Bessel functions $K_0$ and $K_1$, and they depend on the quark electric
charge $e_f=2/3,-1/3$ and mass $m_f$ through
\begin{eqnarray}
\label{eq:psil}
|\Psi_L^{(f)}(r,\al;Q^2) |^2&=&e_f^2\frac{\alpha_e N_c}{2\pi^2} 4Q^2 \al^2(1-\al)^2K_0^2(r \bar{Q}_f)\\
\nonumber
|\Psi_T^{(f)}(r,\al;Q^2) |^2&=&e_f^2\frac{\alpha_e N_c}{2\pi^2}\Bigl( [\al^2+(1-\al)^2]
\bar{Q}_f^2 K_1^2(r \bar{Q}_f)\\
&+& m_f^2 K_0^2(r \bar{Q}_f)\Bigr)
\label{eq:psit}
\end{eqnarray}
for  $\bar{Q}_f^2=\al (1-\al) Q^2+m_f^2$. A common choice is to use $m_u=m_d=m_s=0.14$ GeV, $m_c=1.4$ GeV and $m_b=4.5$ GeV, however, when $m_f=0$, the photon wave functions depend
on $Q$ only through $z=Qr$, a considerable simplification. The electromagnetic fine-structure constant is labeled $\alpha_e$ and
$N_c=3$ is the number of colors in eqs. (\ref{eq:psil}-\ref{eq:psit}).

The dipole cross section in eq. (\ref{eq:conv}) depends on the dipole transverse size $r$ and $\xi$.
Ewerz et al. in Ref. \cite{ewerz} consider $\xi=x$ and
$\xi=W^2$. While theoretically $\xi=W^2$ is more justifiable, the choice of $\xi=x$ allows the dipole model to match experimental
data over a wider range
of $Q^2$ \cite{ewerz}. There are a number of models for the dipole cross section that depend on $x$ rather than $W^2$ \cite{gbw, bgbk,iim,soyez,albacete}. Since our
aim is to invert parametrization of $F_2$ to determine the dipole cross section, our choice of $\xi$ is determined by how $F_2$ is parametrized.
{

Ewerz et al. in Ref. \cite{ewerz} discuss the convolution formula of eq. (\ref{eq:conv}) and consider limiting regimes for the dipole cross section. They have conveniently rewritten the convolution
in terms of smooth functions with properties amenable to treatment with Fourier transforms. First, the integral over $d\al$ and the angular integral
in $d^2r$ in eq. (\ref{eq:conv}) are performed to write
\begin{equation}
\label{eq:f2}
F_2 (\xi,Q^2) =\sum_f  \, Q \int dr h(Qr, m_f r)\frac{1}{r^2}\hat{\sigma}(r,\xi)\ ,
\end{equation}
where
\begin{eqnarray}
\nonumber
&&h(Qr,m_fr) = \frac{Q r^3}{2\pi\alpha_{em}} \int d\al \\
 &&\times \Bigl[ |\Psi_{T}^{(f)}(r,\al;Q^2) |^2 + | \Psi_{L}^{(f)}(r,\al;Q^2)  |^2\Bigr]\ .
\label{eq:hfunc}
\end{eqnarray}
Here, we include the electric charge in the definition of $h$, different from the convention in Ref. \cite{ewerz}.
With $z=Qr$ and further definitions
\begin{eqnarray}
t&=&\ln(Q/Q_0)\\
t'&=&-\ln(r/r_0)
\end{eqnarray}
and defining $z_0\equiv Q_0 r_0$,
\begin{eqnarray}
\nonumber
F_2 (\xi,Q^2) &=&\sum_f \, z_0 e^t
\int dt' h(z_0 e^{t-t'}, m_f r_0 e^{-t'})\\
&\times&
\frac{1}{rr_0}\hat{\sigma}(r,\xi)\mid_{r=r_0\exp(-t')}\ .
\end{eqnarray}
For a single, massless quark of unit charge, a ``toy model,'' this can be rewritten \cite{ewerz}
\begin{eqnarray}
\nonumber
F (\xi,t) &\equiv& F_2(\xi,Q_0^2e^{2t})e^{-t} \\
&=&  \int_{-\infty}^{\infty} dt' \kappa (t-t') S(\xi,t')\ .
\label{eq:toy}
\end{eqnarray}
where
\begin{eqnarray}
S(\xi,t') &=& \frac{1}{rr_0}\hat{\sigma}_0 (r,\xi)\mid _{r=r_0 \exp(-t')}\\
\kappa (\tau) &=& z_0 h(z_0 e^\tau , 0)\ ,
\end{eqnarray}
with $e_f$ in $h(z_0 e^\tau,0)$ set to unity.
The dipole cross section labeled with a subscript ``0'' represents the $N_f=1$, $m_f=0$ case
with $e_f=1$.
Fig. \ref{fig:kappa0} shows $\kappa(\tau)$ versus $\tau$, a function independent of $Q$ for the massless
case where we have set $z_0=2.4$ so that $\kappa(\tau)$ is nearly symmetric around $\tau=0$. Ewerz et al. in Ref. \cite{ewerz} exploit
the peaked nature of $\kappa(\tau)$ at $\tau=0$ to derive approximate relations between $\hat{\sigma}$ and $\partial F_2/\partial t$ for
both the large $Q^2$ and small $Q^2$ regime. Our approach here is to use the smooth, nearly symmetric form of $\kappa(\tau)$ for the massless
quark case to deconvolute the integral in Eq. (\ref{eq:conv}).

\begin{figure}
\centering
 \includegraphics[angle=270,scale=0.35]{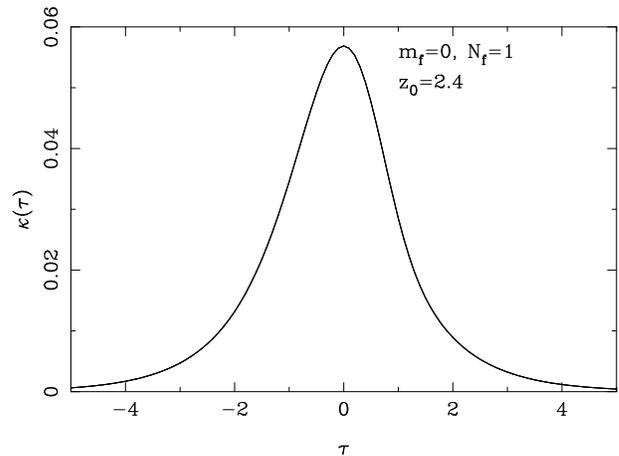}
   \caption{The function $\kappa (\tau)$ as a function of $\tau$ for $m_f=0$ with a single, unit-charge quark flavor. The maximum of $\kappa (\tau)$ is at $\tau=0$ if one chooses $z_0=2.4$, as was done for this figure.}
\label{fig:kappa0}
\end{figure}

Given $F(\xi,t)$ in Eq. (\ref{eq:toy}) for a wide range of $t$ (namely, a wide range of $Q$),  $S(\xi,t')$ can be found by
taking the Fourier transform to factorize the convolution, then inverting the Fourier transform.
In the next section, we perform this procedure numerically where we use
$F_2(x,Q^2)$ parametrized by Donnachie and Landshoff ($\xi=x$) \cite{dl}.
Even though this parametrization has a limited range of applicability  in its description of the
experimental data (e.g., 0.1 GeV$^2<Q^2<10$ GeV$^2$), we use this simple form to demonstrate the procedure and to interpret the results for $\hat{\sigma}$. 
In Sec. IV, for massive quarks, we show results for the Donnachie-Landshoff and Block et al. \cite{blocknew} parametrization.

\section{Massless, unit charge quark example}

As a starting point for using the Fourier transform to extract $\hat{\sigma}$ from a parametrization of $F_2$, we consider
the toy model of a massless quark with unit charge.

Using the usual formulas for the Fourier transform and its inverse,
\begin{eqnarray}
f(t)&=&\int_0^\infty dk\{ a_f(k)\cos kt + b_f(k) \sin kt\}\\
a_f(k) &=& \frac{1}{\pi}\int_{-\infty}^{\infty}dt \cos kt\  f(t)\\
b_f(k) &=& \frac{1}{\pi}\int_{-\infty}^{\infty}dt \sin kt \ f(t)\ ,
\end{eqnarray}
for a completely symmetric function $\kappa(\tau)$, $b_\kappa (k)=0$.
We approximate $b_\kappa (k)\simeq 0$ in what follows.
We find that the dipole cross section we obtain, when convoluted with the photon wave function and appropriately normalized, is a good
approximation to the
parametrized structure function.

Because in the massless case $\kappa(\tau)$ depends only on $\tau$, the convolution integral factorizes with Fourier transformed and simplifies to
the following when $b_\kappa=0$:
\begin{eqnarray*}
\pi a_F(k) &=& \int_{-\infty}^{\infty} dt \cos kt \int_{-\infty}^{\infty} dt' \kappa(t-t') S(t')\\
&=& \int_{-\infty}^{\infty} d\tau dt' \cos k\tau\, \cos kt'\, \kappa (\tau) S(t')\\
&=& \pi a_\kappa (k)\, \pi a_S(k)
\end{eqnarray*}
and
\begin{eqnarray*}
\pi b_F(k) &=& \int_{-\infty}^{\infty} dt \sin kt \int_{-\infty}^{\infty} dt' \kappa(t-t') S(t')\\
&=& \int_{-\infty}^{\infty} d\tau dt' \cos k\tau\, \sin kt'\, \kappa (\tau) S(t')\\
&=& \pi a_\kappa (k)\, \pi b_S(k) \ .
\end{eqnarray*}
The Fourier transform of $\kappa$ is straightforward. For the Fourier transform of $F$, as noted above, we need $F_2(x,Q^2=Q_0^2e^{2t})$
for the full range of $Q^2$, in principle between $Q^2=0\to \infty$. The Donnachie-Landshoff parametrization \cite{dl} meets the requirement of
being defined for all $Q^2$, even if not valid for the full range. The full expression for $F_2(x,Q^2)$ is
listed in Appendix A, eq. (\ref{eq:f2dl}). For small $x$,
\begin{equation}
F_2(x,Q^2)\sim Ax^{1-\alpha}\Biggl(\frac{Q^2}{Q^2+a}\Biggr)^\alpha
 \,
\label{eq:f2dlapprox}
\end{equation}
where $\alpha=1.0808$.
Our comparison will be restricted to the $Q^2$ range between $0.1-10$ GeV$^2$ where the D-L fit was performed.
We have chosen $Q_0=0.82$ GeV so that $F(x,t)$ has a maximum at $t=0$. While it is not completely symmetric, $F(x,t)$ has a similar shape for
$t$ less than and greater than zero, for small $x$.

\begingroup
\squeezetable
\begin{table}[h]
\begin{center}
\vskip 0.25in
\begin{tabular}{|c|c|c|c|c|c|}
\hline
 r [GeV$^{-1}$] & a & b & c & d &e \\
\hline
\hline
$r < 1$ & 0.02315 & 3.34 &  &  &   \\
\hline
$1 \leq r < 2.9$ & 0.9943 & $7.259 \times 10^{-3}$ & 4.3028 & $2.469 \times 10^{-2}$ & 4.0702 \\
\hline
$2.9 \leq r < 5 $ & 2.7866 & $-1.762 \times 10^{-2}$ & 4.1789 & 6.4517 & -0.8983  \\
\hline
$r \geq 5$ & 35.2489 & -2.62 &  &  &   \\
\hline
\hline
\end{tabular}
\caption{\label{table:param}
The values of the parameter in the dipole cross section formula, with $s_0=27.95$ mb and $r$ is in units of GeV$^{-1}$.}\end{center}
\vspace{-0.6cm}
\end{table}
\endgroup

The resulting dipole cross section $\hat{\sigma}_0$ from the Fourier transform and inversion of $F$ and $\kappa$, for $m_f=0$ with a unit
charged quark, is shown in
Fig. \ref{fig:hatsig0}. Three values of $x$ are shown, $x=10^{-3},\ 10^{-4}$ and $10^{-5}$ in increasing order.
The dipole cross section can be parametrized as
\begin{eqnarray}
\label{eq:sig0}
\hat{\sigma}_0 &=& s_0 \biggl(\frac{x}{10^{-4}}\biggr)^{1-\alpha}
(a r^b),\\
\nonumber
& & {\rm for}\ r < 1\ {\rm GeV^{-1}},\ r\geq 5 \ {\rm GeV^{-1}} \\
\hat{\sigma}_0 &=& s_0\biggl(\frac{x}{10^{-4}}\biggr)^{1-\alpha}
(\pm 1-a e^{(b r^c)}+d r^e),\\
\nonumber
& & + \ {\rm for} \ 1 \ {\rm GeV^{-1}}\leq r < 2.9 \ {\rm GeV^{-1}}\\
\nonumber
& & - \ {\rm for} \ 2.9\ {\rm GeV^{-1}} \leq r < 5\ {\rm GeV^{-1}}
\end{eqnarray}
where $s_0$ is 27.95 mb and $r$ is in units of GeV$^{-1}$.
The other parameters are presented in Table \ref{table:param}.
The $x$ dependence
follows directly from the $x$ dependence of $F_2$ in the Donnachie-Landshoff parametrization, where for this range of $x$ and $Q^2>0.1$
GeV$^2$, the first term in Eq. (\ref{eq:f2dl}) dominates.
The rising behavior at low $r$ is characteristic of other parametrization of the dipole cross section, but the falling behavior is not.
We compare our dipole form to theoretically motivated $\hat{\sigma}$ in Sec. IV.

\begin{figure}
\centering
\includegraphics[angle=270,scale=0.35]{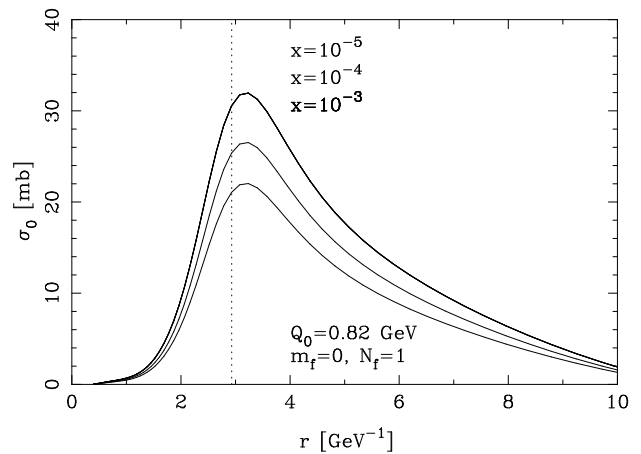}
   \caption{The quantity ${\hat{\sigma}}_0$ as a function of $r$ given a unit-charged, massless quark, for $x=10^{-5},\ x=10^{-4}$ and $x=10^{-3}$ using the Donnachie-Landshoff parametrization of $F_2$. The vertical dotted line shows $r_0=2.93$ GeV$^{-1}$=0.58 fm. }
\label{fig:hatsig0}
\end{figure}

\begin{figure}
\centering
\includegraphics[scale=0.35]{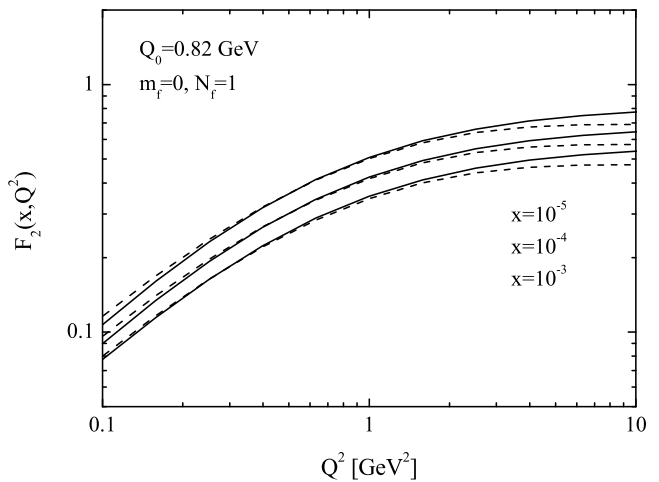}
   \caption{The resulting $F_2(x,Q^2)$ as a function of $Q^2$ for $x=10^{-5},\ 10^{-4},\ 10^{-3}$
using the dipole cross section
shown in Fig. 2 (dashed) and the Donnachie-Landshoff parametrization for Eq. (\ref{eq:f2dl}).}
\label{fig:f20}
\end{figure}

Fig. \ref{fig:f20} shows the result for $F_2$ when the dipole cross section shown in Fig. \ref{fig:hatsig0} is convoluted with the massless unit-charged
quark wave function. It matches well with the original parametrization up to $Q\simeq 2-3$ GeV. At $Q^2=0.1$ GeV$^2$,
the dipole formula with $\hat{\sigma}_0$ is about { 7\%} 
larger than the Donnachie-Landshoff parametrization of $F_2$,
while for $Q^2=10$ GeV$^2$, the dipole formula is about { 11\%} 
below the D-L parametrization.
We do not get an exact match to $F_2$ in our inversion because we have made the approximation $b_\kappa(k)=0$ and some numerical approximations in our integration at large $k$.
Nevertheless, the procedure
works reasonably well. As we show below, when $m_f\neq 0$, this approximate $\sigma_0$ is sufficient.

\section{Massive quarks}

\begin{figure}
\centering
 \includegraphics[angle=270,scale=0.35]{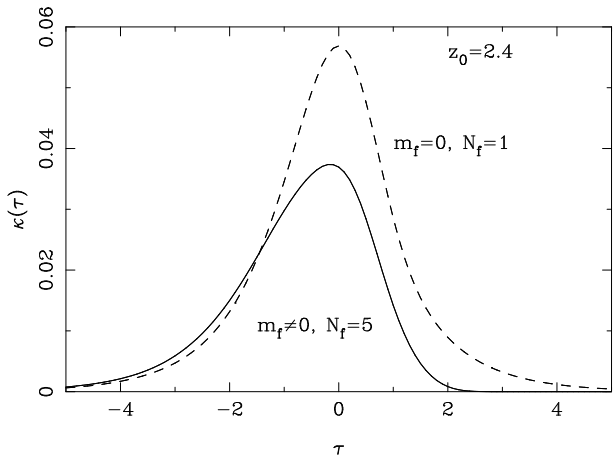}
   \caption{The function $\kappa (\tau)$ as a function of $\tau$ for five quark flavors ($m_f\neq 0$) and $Q=1$ GeV shown with the solid line, and
for $m_f=0$, as in Fig. \ref{fig:kappa0} with one flavor shown with the dashed line.}
\label{fig:kappa}
\end{figure}

For massive quarks, eq. (\ref{eq:conv}) involves a sum over flavors and the introduction of an $r$ dependence in $h(z=Qr,m_fr)$. 
This means that the factorization of eq. (\ref{eq:toy}) via the Fourier transform is not exact. The dipole cross section from the toy model, appropriately normalized, is nevertheless a reasonably good approximation to the solution in the massive case. We cannot exactly solve the inversion problem for the massive case, but we show in this section how our approximate dipole cross section nearly reproduces $F_2(x,Q^2)$ using eq. (\ref{eq:f2}).

 Rewriting $m_f r=m_f z/Q=m_f z_0e^\tau/Q $ in eq. (\ref{eq:hfunc}), we express $\kappa$ for the massive case as 
\begin{equation}
\kappa(\tau, m_f z/Q) = \sum_{f} z_0 h(z_0 e^{\tau}, m_f z_0e^\tau/Q ) \ .
\label{eq:kapmf}
\end{equation}
Fig. \ref{fig:kappa} shows this $\kappa$ for massive quarks for five flavors with $Q=1$ GeV. For comparison, $\kappa$ for $m_f=0$ is also presented with the dashed line. 
To find $\hat{\sigma}$, we take as our starting point $\hat{\sigma}_0$. Since 
the $\gamma^* p$ center of mass energy squared $W^2\simeq Q^2/x$, for $x<5 \times 10^{-3} $ we have $W^2>m_b^2$, so we use all five flavors in the massive case in all the examples shown below.

The toy model $\hat{\sigma}_0$ must be normalized to account for the
electric charge and the sum over flavors with massive quarks. 
For massive quarks, we show the results for both the Donnachie-Landshoff \cite{dl} and the Block et al. \cite{blocknew} parametrization
of $F_2(x,Q^2)$. For the Donnachie-Landshoff parametrization, $N_\sigma$ is chosen by matching $F_2$ calculated with the normalized dipole cross section to $F_2(x,Q^2=1\ {\rm GeV}^2)$. The value of $N_\sigma$ for the Block et al. parametrization is chosen for each $\xi$  value in the range of $Q^2=1.6-2.5$ GeV$^2$. The specific value of $Q^2$ for the normalization is picked to minimize the discrepancy between the $F_2$ calculated with the dipole approach using eq. (5) and the initial parametrization of $F_2$.

\subsection{Donnachie-Landshoff $F_2$}

In this section, taking the Donnachie-Landshoff $F_2(x,Q^2)$,
we find
\begin{equation}
\label{eq:nsig}
\hat{\sigma} = 1.57 \hat{\sigma}_0 = N_\sigma \hat{\sigma}_0\ .
\end{equation}
The factor $N_{\sigma}$ accounts for the effect of the charge and the masses of quarks.
{
We find that for the Donnachie-Landshoff parametrization of $F_2$, $N_\sigma$ does not depend on $x$ values. 


With the inclusion of masses for five quark flavors and $N_\sigma$, we show  $F_2(x,Q^2)$ versus $Q^2$ for
$x=10^{-3},10^{-4},10^{-5}$ in Fig. \ref{fig:f25}. The agreement between the original Donnachie-Landshoff parametrization shown
by the solid lines and the structure function evaluated using $\hat{\sigma}$ is quite good for the range of validity of the DL parametrization.

\begin{figure}
\centering
\includegraphics[scale=0.35]{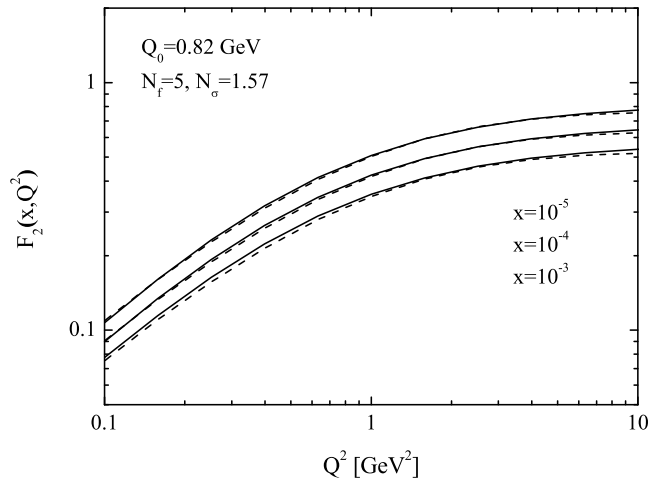}
   \caption{The resulting $F_2(x,Q^2)$ as a function of $Q^2$ for $x=10^{-5},\ 10^{-4},\ 10^{-3}$
using the dipole cross section $\hat{\sigma}_0$
shown in Fig. \ref{fig:hatsig0} multiplied by $N_\sigma = 1.57$ (dashed) and the Donnachie-Landshoff parametrization for eq. (\ref{eq:f2dl}) }
\label{fig:f25}
\end{figure}

For reference, we illustrate the range of $r$ most relevant for the evaluation of $F_2$. 
Fig. \ref{fig:f2frac} shows the ratio
\begin{equation}
f(r_{max}) = \frac{F_2(x,Q^2,r_{max})}{F_2(x,Q^2)} \ ,
\end{equation}
where $r_{max}$ is the upper limit of the r-integration in eq. (\ref{eq:conv}). 
The function $F_2(x,Q^2)$ in the denominator is evaluated with $r_{max} \to \infty$. 

When $Q^2=10$ GeV$^2$, approximately 10\% of the structure
function comes from low $\hat{\sigma}$ at low $r$, $r\lsim 1$ GeV$^{-1}$, and approximately 10\% comes from $r\gsim 4$ GeV$^{-1}$.
For $Q^2=0.1$ GeV$^2$, 80\% of the evaluation of $F_2(x,Q^2)$ comes from $r\sim 2-{5}$ GeV$^{-1}$.
Thus, the bulk of the evaluation of $F_2$ for the range of $Q^2$ of interest
for the Donnachie-Landshoff parametrization is in the range of $r\sim 1-{5}$ GeV$^{-1}$
where the dipole cross section makes the transition from the low $r$ to large $r$ forms.

\begin{figure}
\centering
\includegraphics[scale=0.35]{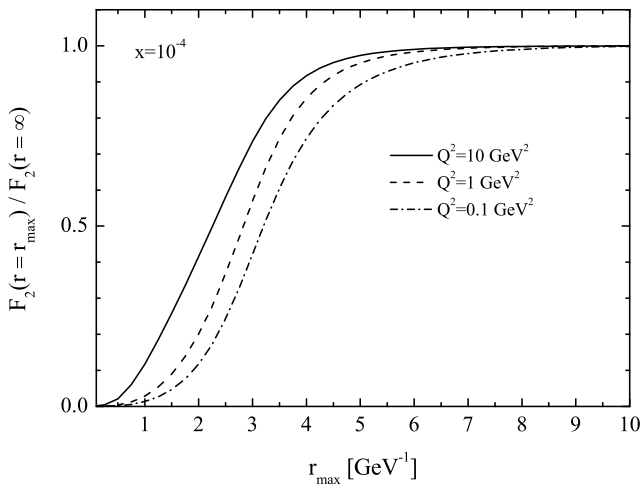}
   \caption{The fraction $f=F_2(x,Q^2,r_{max})/F_2(x,Q^2,r_{max}\to \infty)$ as a function of $r_{max}$ for $x=10^{-4}$ for
$Q^2=0.1,1,10$ GeV$^2$ from lowest to highest curve.
}
\label{fig:f2frac}
\end{figure}

\begin{figure}
\centering
\includegraphics[scale=0.35,angle=270]{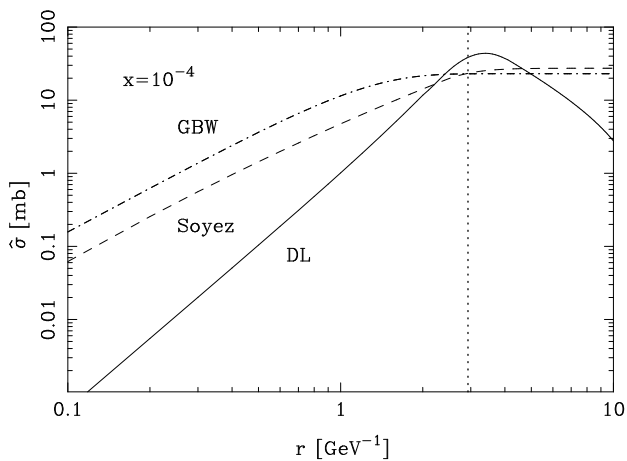}
   \caption{The dipole cross section  approximated here via Fourier transform inversion and rescaling of the
Donnachie Landshoff (D-L) $F_2$ (solid) \cite{dl}, the Soyez parametrization (dashed) \cite{soyez} and
Golec-Biernat and W\"usthoff (GBW) dipole formula (dot-dashed) \cite{gbw}
for $x=10^{-4}$.  The vertical dotted line shows $r=r_0=2.93$ GeV$^{-1}$. }
\label{fig:sdcompare}
\end{figure}


The low $r$ and large $r$ approximate forms of the dipole cross section and their
relation to $F_2$ are discussed in Ref. \cite{ewerz}. Our dipole cross section does not precisely match
these limiting forms.
For small $r$, Ewerz et al. have shown that \cite{ewerz},
\begin{eqnarray}
\nonumber
\hat{\sigma}(r,\xi)&\simeq& {\pi^3} r^2 Q^2\frac{\partial}{\partial Q^2}F_2(\xi,Q^2)\mid_{Q^2=(z_0/r)^2}\ ,
\\
\label{eq:emnlow}
&& {\rm for}\  r\ll 0.3\ {\rm GeV^{-1}} .
\end{eqnarray}
This relation relies on approximations including that $\hat{\sigma}(r,x)/r^2$ is slowly varying
for small $r$ and on the approximate relation that $Qr\simeq z_0$.
At large $Q^2$ (equivalent to small $r$), $Q^2\partial F_2/\partial Q^2
\sim 1/Q^2$, so $\hat{\sigma}\sim r^2/Q^2\sim r^4/z_0^2$. 
The dipole cross section we obtain scales approximately as $\hat{\sigma}\sim r^{3.34}$ around $r\sim
1$ GeV$^{-1}$. Given that $\hat{\sigma}(r,x)/r^2$ is not slowly varying for small $r$, one would not
expect a precise agreement between our low $r$ parametrization and Eq. (\ref{eq:emnlow}).

For large $r$, the small $Q^2$ regime, the approximate behavior of $\hat{\sigma}$ is related to the
logarithmic derivative of the virtual photon cross section. To get the correct physical behavior
of $\sigma_{\gamma^* p}$ for $Q^2\to 0$, one needs to fix $W^2\simeq Q^2/x$
in $\sigma _{\gamma^* p}\propto F_2/Q^2$. We therefore
write,
\begin{eqnarray}
\nonumber
\hat{\sigma}(r,x)&\simeq&
 -\frac{\pi}{\alpha_{e}} Q^2
\frac{\partial}{\partial Q^2}
\Bigl[ \sigma_T(W^2,Q^2) + \sigma_L(W^2,Q^2)\Bigr]\\
\label{eq:emnhi}
&& {\rm with\ } {Q^2=(z_0/r)^2,\ W^2\ {\rm fixed}} .
\end{eqnarray}
The Donnachie-Landshoff parametrization of $F_2$ together with this large $r$
approximation leads to $\hat{\sigma}\sim r^{-2}$. The dipole cross section we determined
has $\hat{\sigma}\sim r^{-2.62}$ for $r\simeq 5$ GeV$^{-1}$.
Eq. (\ref{eq:emnhi}) also relies on $\hat{\sigma}$ being a slowly varying function of $r$ at large $r$, so the fact that our approximate power law behavior deviates from $\hat{\sigma}\sim
r^{-2}$ is again not surprising.  As remarked in Ref. \cite{ewerz}, the falling
dipole cross section as a function of large $r$ is correct when one uses the standard perturbative photon wave function for all $Q^2$, even for very low $Q^2$. 

 In Fig. 7, we compare $\hat{\sigma}$ with two examples of dipole cross sections, the Golec-Biernat and W\"usthoff (GBW) cross section \cite{gbw} and the Soyez parametrization \cite{soyez} of the dipole cross section based on the Iancu-Itakura-Munier form \cite{iim}. Explicit expressions are included in Appendix C. Given the limited applicability of the D-L parametrization, it is not surprising that the 
GBW and Soyez dipole cross sections do not match the solid line.
The low $r$ form of the GBW dipole scales as $\hat{\sigma}\sim r^2$ as $r\to 0$, and
the GBW dipole has
$\hat{\sigma}\to \sigma_c$, a constant value, for large $r$.
The Soyez dipole in Ref. \cite{soyez} scales as $\hat{\sigma}\sim r^{1.7}$ for $r=0.3$ GeV$^{-1}$ and $x=10^{-4}$. The large $r$ form for this
dipole formula goes to a similar constant value.

Our approach here is to use the standard perturbative wave function even at the lowest $Q^2$ values. The result is that we find that
our approximate $\hat{\sigma}(r,x)$ decreases at large $r$, a feature pointed out in Ref. \cite{ewerz}. While the D-L structure function is not applicable to a wide range of $Q^2$, we also find a similar 
large $r$ behavior in the next section, using the Block et al.
parametrization.   As noted in Ref. \cite{ewerz},
with the standard perturbative photon wave function,
the GBW and Soyez forms of the dipole cross section yield a virtual photon-proton cross section $\sigma_T(W^2,Q^2) + \sigma_L(W^2,Q^2)$ that
does not have the correct $Q^2\to 0$ behavior without a
modification of the low $Q^2$ photon wave function.

\subsection{Block et al. $F_2$}

The Donnachie-Landshoff parametrization of $F_2$ has a limited range of applicability.  
More recent parametrization of $F_2$ that apply to large and small $Q^2$ and small $x$ have been presented in, for example, Refs. \cite{haidt0,block1,block2,block3,
block4,block5,haidt,block,blocknew}.
In particular, the new parametrization of Block, Durand and Ha in Ref. \cite{blocknew} has an expression for the asymptotic part of $F_2$ (no-valence) that  accounts for the asymptotic behavior ($W^2\to\infty$ with $Q^2$ fixed),
\begin{equation}
F_2(W^2,Q^2)\propto \ln^2(W^2/Q^2)\simeq \ln^2(1/x)
\end{equation}
for small $x$. It also has the feature we need for the Fourier inversion: a well defined parametrization for the full range of $Q^2$.
The parametrization appears in Appendix B, eq. (\ref{eq:blocknewx}) in terms of $x$ and
eq. (\ref{eq:blocknewW}) in terms of $W^2$.
The two forms of $F_2(x,Q^2)$ permit the evaluation of $\hat{\sigma}$ with $\xi=x$ and $\xi=W^2$. 

We begin by setting $\xi=x$, then finding $\sigma_0$ and $N_\sigma(x)$. Again, our resulting 
$\hat{\sigma} =N_\sigma(x)\hat{\sigma}_0(r,x)$ is only an approximate result from the combination of the approximate solution to the massless, unit charge case multiplied by a normalization factor, here dependent on $x$. Sample values of the normalization factors are $N_\sigma(x=10^{-3})=1.40$ and
$N_\sigma(x=10^{-5})=1.25$.
The dipole cross section for 
$x=10^{-3}, \ 10^{-4}$ and $10^{-5}$, in ascending order, is shown with the solid curves in Fig. \ref{fig:hatsigx}. 
{ The double peaked form of the dipole cross section appears from the sum of the three terms in eq. (\ref{eq:blocknewx}).}

Fig.  \ref{fig:hatsigx} also shows the dipole cross sections of Soyez \cite{soyez} (dot-dashed line) and the low $r$ approximation by Ewerz et al. \cite{ewerz} shown in eq. (\ref{eq:sigmadip2}) and eq. (\ref{eq:emnlow}), respectively. 
The approximate power law at low $r$ for the solid line is 
\begin{equation}
\hat{\sigma} \sim r^{1.9}\quad {\rm for}\ r \lsim 0.5\ {\rm GeV}^{-1}\ ,
\end{equation}
which is fairly well matched to the Soyez, GBW and low $r$ functional forms of the dipole cross section.

The resulting
structure functions $F_2(x,Q^2)$ for these $x$ values are shown in Fig. \ref{fig:f25blockx}. The dashed lines show the calculation of $F_2$ with the dipole formula and the solid lines are directly the parametrization of Block et al., eq. (\ref{eq:blocknewx}). 
The difference between these two structure functions  in the range shown here is the biggest at $Q^2=10\ {\rm GeV^2}$, with a disagreement at the level of about $10 \%$ for $x \leq 10^{-4}$.
For other $Q^2$ values in the range, the agreement to the original parametrization of $F_2(x,Q^2)$ is better as $x$ becomes smaller.


\begin{figure}
\centering
\includegraphics[scale=0.35]{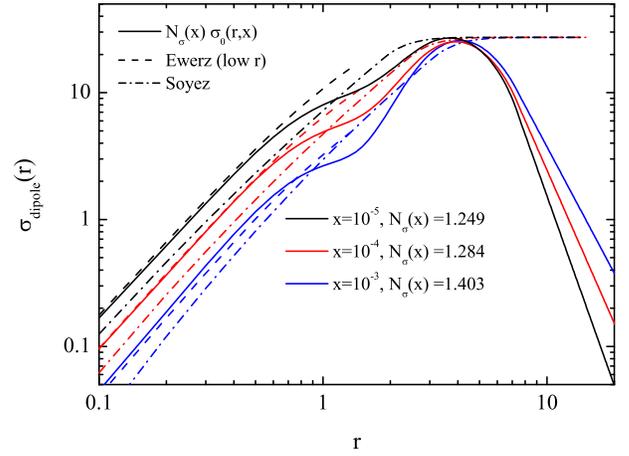}
   \caption{The quantity ${\hat{\sigma}}=N_\sigma(x)\hat{\sigma}_0(r,x)$ as a function of $r$  for $x=10^{-5},\ x=10^{-4}$ and $x=10^{-3}$ using the Block et al. parametrization of $F_2(x,Q^2)$, shown by the solid lines. Also shown is the Soyez dipole of eq. (\ref{eq:sigmadip2}) (dot-dashed) and the small $r$ extrapolation of Ewerz et al. in eq. (\ref{eq:emnlow}) with $\xi=x$ held fixed (dashed).}
\label{fig:hatsigx}
\end{figure}

\begin{figure}
\centering
\includegraphics[scale=0.35]{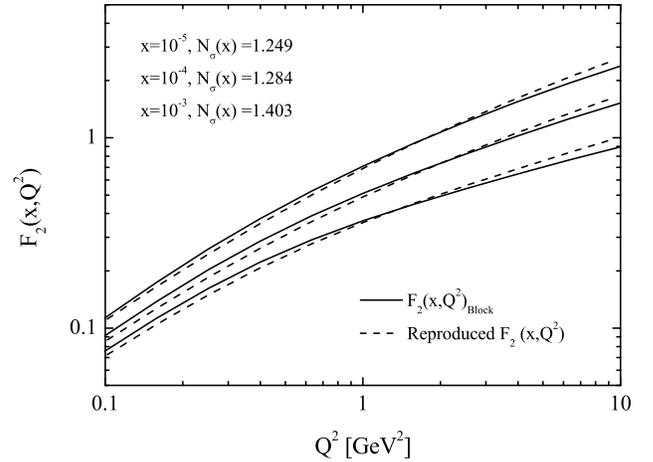}
   \caption{The resulting $F_2(x,Q^2)$ as a function of $Q^2$ for $x=10^{-5},\ 10^{-4},\ 10^{-3}$
using the dipole cross section $\hat{\sigma}=N_\sigma \hat{\sigma}_0$ (dashed) and the Block et al. parametrization of eq. (\ref{eq:blocknewx}) (solid lines). }
\label{fig:f25blockx}
\end{figure}

\begin{figure}
\centering
\includegraphics[scale=0.35]{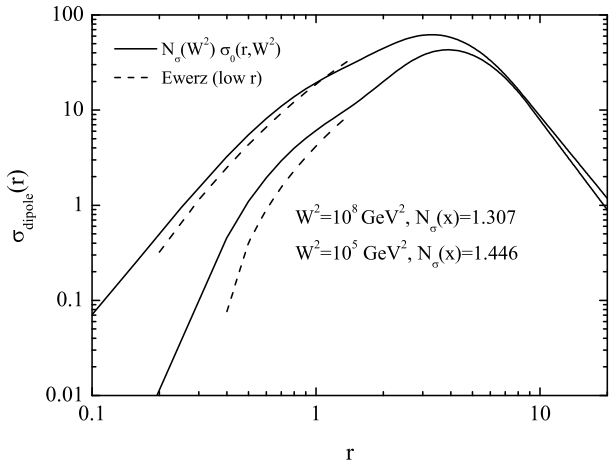}
   \caption{The quantity ${\hat{\sigma}}=N_\sigma(W^2)\sigma_0(r,W^2)$ as a function of $r$  for $W^2=10^{5}$ GeV$^2$ and $W=10^{8}$ 
GeV$^2$ using the Block et al. parametrization of $F_2(W^2,Q^2)$. Also shown is the small $r$ extrapolation of Ewerz et al. in eq. (\ref{eq:emnlow}) with $\xi=W^2$ held fixed (dashed).}
\label{fig:hatsigw}
\end{figure}

\begin{figure}
\centering
\includegraphics[scale=0.35]{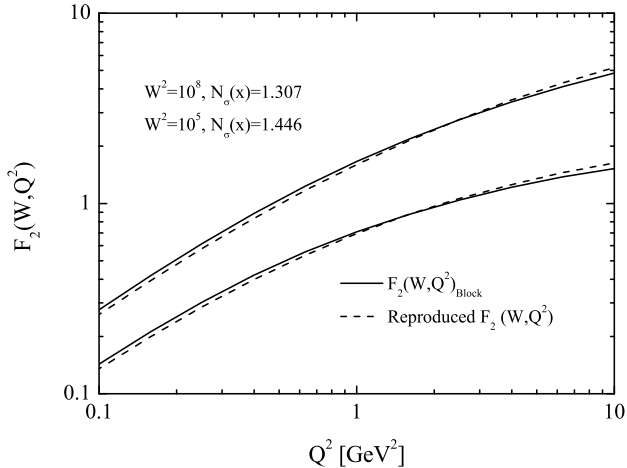}
   \caption{The resulting $F_2(W^2,Q^2)$ as a function of $Q^2$ for $W^2=10^{5},\ 10^{8}$ GeV$^2$
using the dipole cross section $\hat{\sigma}=N_\sigma \hat{\sigma}_0$ (dashed) and the Block et al. parametrization of eq. (\ref{eq:blocknewW}). }
\label{fig:f25blockw}
\end{figure}

Figs. \ref{fig:hatsigw} and \ref{fig:f25blockw} show the dipole cross section $\hat{\sigma}(r,W^2)$ and
$F_2(W^2,Q^2)$ respectively. The integral of $F(t)$ with $W^2$ held fixed has a different dependence on $Q^2$ (and therefore $t$ dependence) than the integral with $x$ held fixed. When $\xi=W^2$ is held fixed, the resulting
dipole cross section does not have the double peak. The low $r$ approximate form for $\hat{\sigma}$ following Ref. \cite{ewerz} does not match
as well for $\xi=W^2$ as for $\xi=x$, and it drops rapidly as $r\to 0$. Indeed, $\hat{\sigma}$ becomes negative for non-zero values of $r$. This unphysical behavior has been noted in Ref. \cite{ewerz,nachtmann1,nachtmann2,nachtmann3}, an indication that a large, fixed $W^2$, the dipole
approach loses its validity at large $Q^2\simeq z_0^2/r^2$. To avoid a negative dipole cross section, we extrapolate $\hat{\sigma}(r,W^2)$ with
a power law set at $r=0.3$ GeV$^{-1}$. 

The ad hoc extrapolation of $\hat{\sigma}(r,W^2)$ at low $r$ nevertheless allows us to reproduce $F_2$ reasonably well in the $Q$ range of interest, $0.1$ GeV$^2<Q^2<10$ GeV$^2$. 
Using the dipole cross section in Fig. \ref{fig:hatsigw}, the calculated $F_2$ is about 5\% lower than the Block et al. parametrization for $Q^2=0.1$ GeV$^2$, and about 7\% higher for $Q^2$=10 GeV$^2$, for both $W^2=10^5$ and $10^8$ GeV$^2$.

We have also used the relation between $x$ and $W^2$,
\begin{equation}
x=\frac{Q^2}{Q^2+W^2-m_p^2}\ 
\end{equation}
and $\hat{\sigma}(r,W^2)$ to find 
$F_2(x,Q^2)$. We find that for $x=10^{-3}-10^{-7}$, at $Q^2\simeq 0.01$ GeV$^2$, we can reproduce $F_2(x,Q^2)$ to better than 1\% using
$\hat{\sigma}(r,W^2)$. Using $\hat{\sigma}(r,x)$, as $x$ decreases, the agreement between the dipole calculation and the Block et al. parametrization is not as good. For $x=10^{-7}$ and $Q^2=0.01$ GeV$^2$, using $\hat{\sigma}(r,x)$ differs from the 
Block et al. parametrized $F_2$ by approximately 20\%.

\begin{figure}
\centering
\includegraphics[scale=0.33]{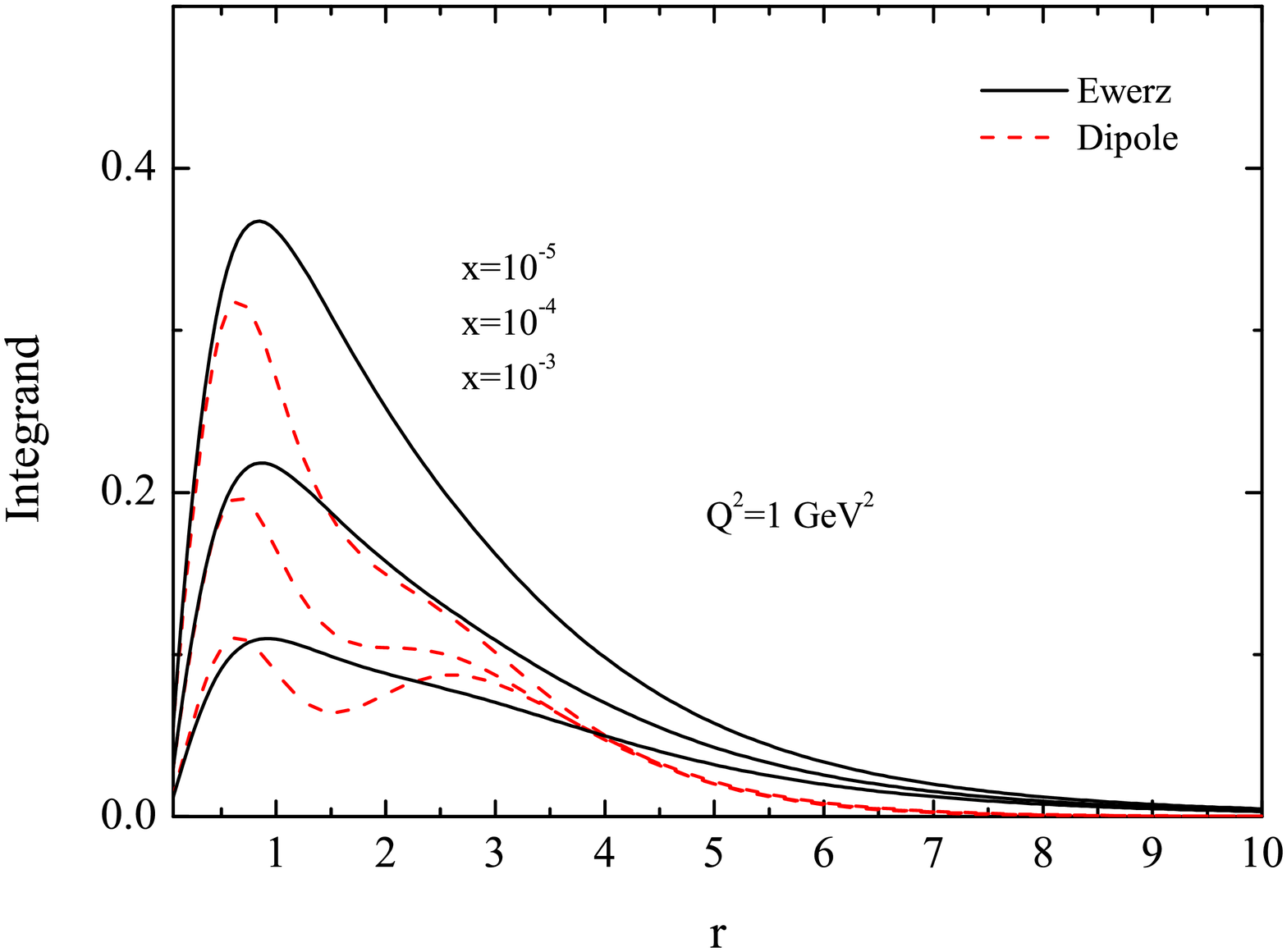}
\includegraphics[scale=0.33]{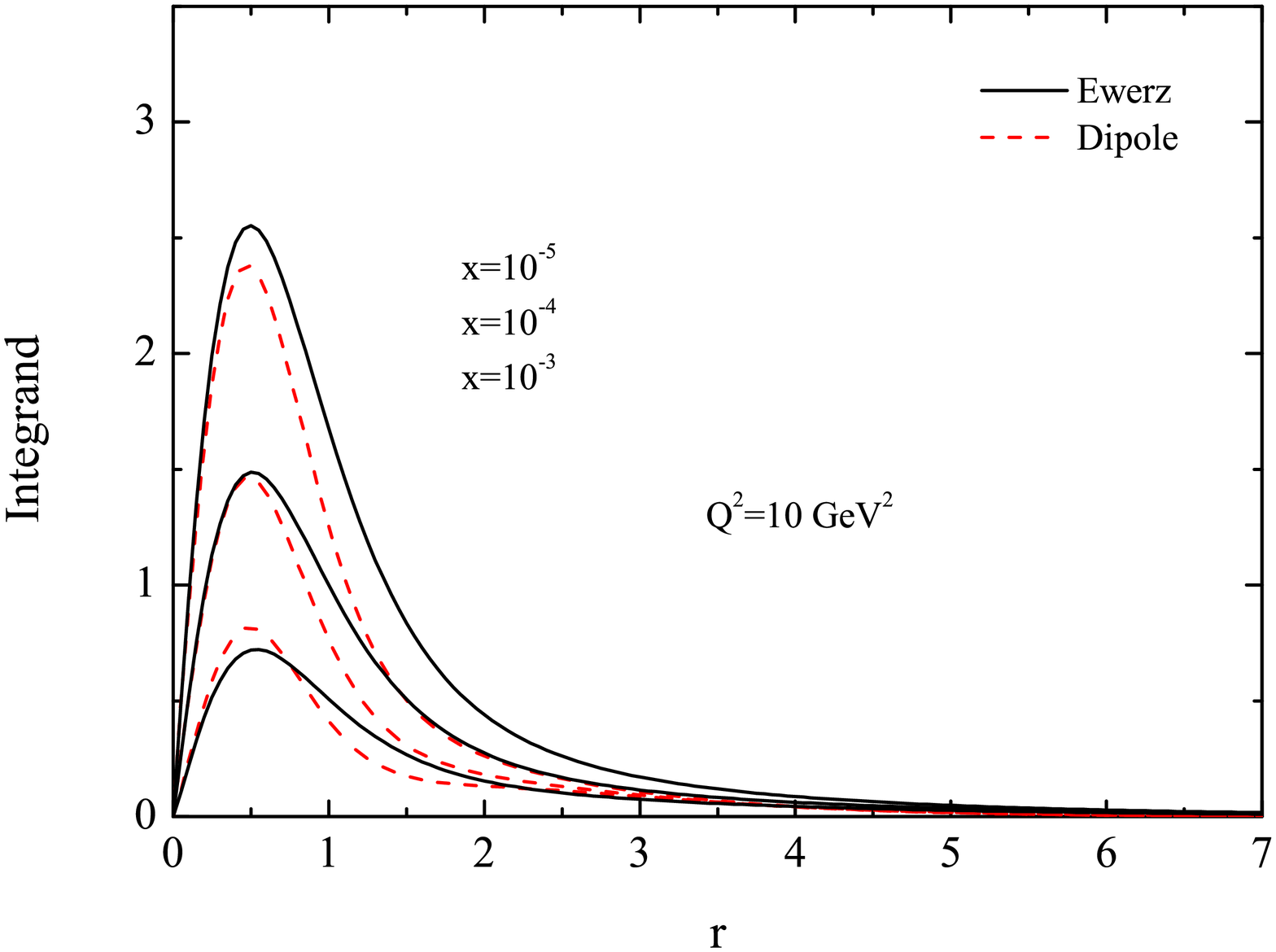}
\includegraphics[scale=0.33]{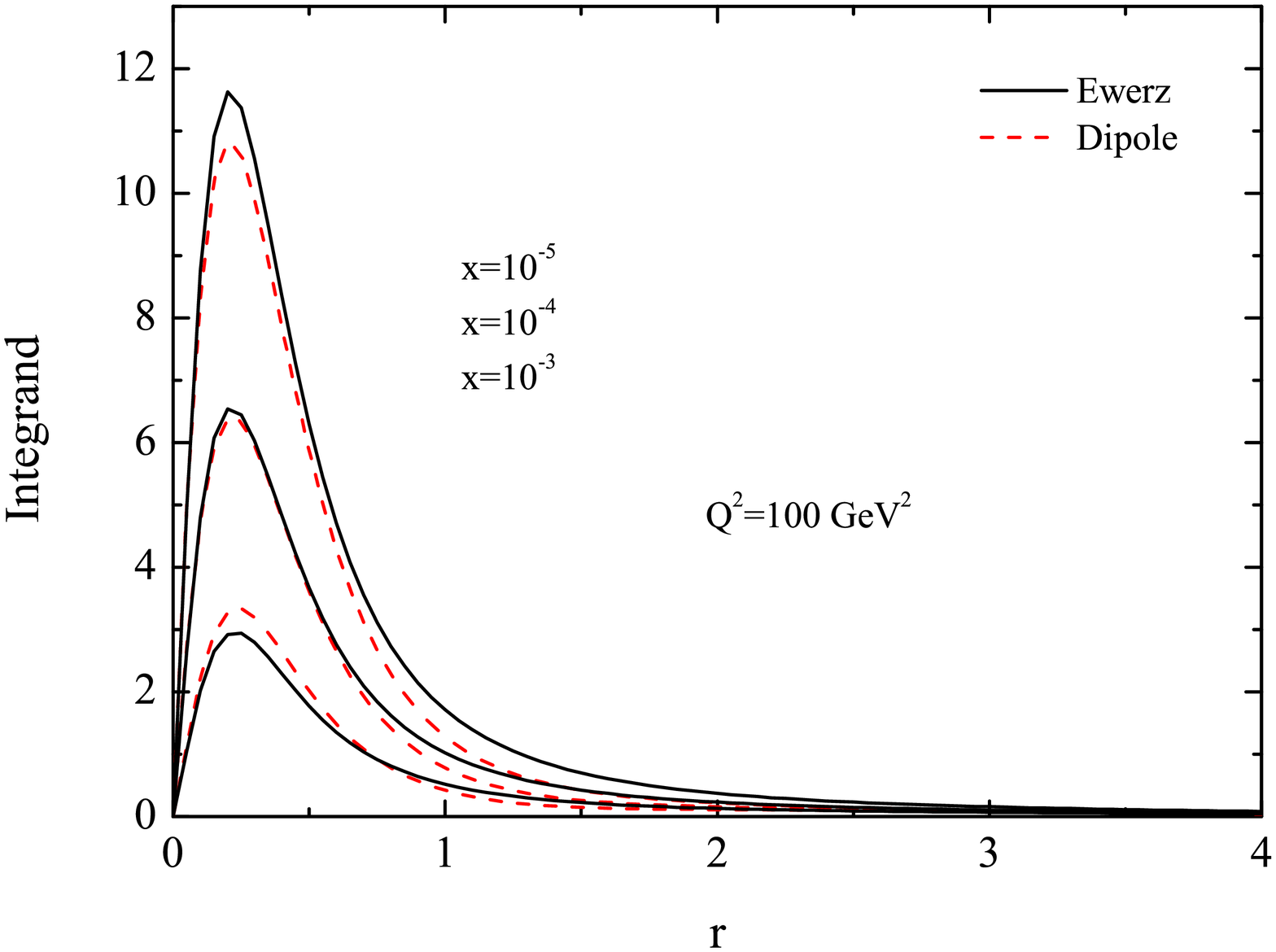}
  \caption{The integrand in eq. (5) times $Q$ to determine $F_2$ for $\xi=x=10^{-3} ,\ 10^{-4},\ 10^{-5}$ (from lowest to highest curve)
and $Q^2=1,\ 10, 100$ GeV$^2$. The dashed curves employ $\hat{\sigma}=N_\sigma \hat{\sigma}_0$ approximated from the Block et al.
parametrization, while the solid curves use eq. (\ref{eq:emnlow}).
 }
\label{fig:integrand}
\end{figure}

Finally, we show in Fig. \ref{fig:integrand} plots of the integrand in eq. (5). The dashed curves show
$\hat{\sigma}$ found here, multiplied by $\sum_f Q h(Qr,m_f r)/r^2$, while the solid curves use instead the approximate expression for $\hat{\sigma}$ from eq. (\ref{eq:emnlow}) where $\xi=x$. The low $r$ shape of the integrand
is well approximated by eq. (\ref{eq:emnlow}), however, these figures show that for the $Q^2$ values around 1 GeV$^2$, the detailed shapes
do not agree very well.  As
$Q^2$ 
increases, the approximate formula for the small $r$ multiplied by  $\sum_f Q h(Qr,m_f r)/r^2$
better represents the integrand.  With suitable normalization, the approximately formula for $\hat{\sigma}$ from eq. (\ref{eq:emnlow}) may be sufficient to represent the dipole cross section.

\section{Conclusions}

In summary, we have used a simplified form of the convolution formula in the color dipole formalism
with $m_f=0$, and rescaled
to find an approximate dipole cross section from a parametrization of the structure function $F_2(x,Q^2)$.
We have used the Donnachie-Landshoff \cite{dl}  and Block et al. \cite{blocknew} parametrization to illustrate that this procedure approximates the starting $F_2(x,Q^2)$ to within about
10\% for small $x$ and $Q^2=0.1-10$ GeV$^2$. 

For the Block et al. parametrization of $F_2$, we are able to consider both $\xi=x$ and $\xi=W^2$. For $Q^2=0.1-10$ GeV$^2$,
there doesn't seem to be an advantage to one particular choice of $\xi$. For lower $Q^2$, $Q^2\sim 0.01$ GeV$^2$, using
$\xi=W^2$ to find $\hat{\sigma}$ better reproduces 
$F_2$, while for larger $Q^2$, $\xi=x$ better reproduces $F_2$, although neither choice works at the 10\% level for $Q^2=100$ GeV$^2$.
For 
$Q^2\simeq 100$ GeV, the errors are between $\sim 20-30\%$ for $x=10^{-3}-10^{-5}$.
The errors are larger when $x\simeq 10^{-3}$ than for smaller values of $x$. Smaller $x$ means
well away from the valence regime, and $W^2\simeq Q^2/x \gg (2m_b)^2$, the threshold for $b$ quark pair production.

We also consider the small and large $r$ behavior of $\hat{\sigma}$, following the 
discussion in Ref. \cite{ewerz}.
The Donnachie-Landshoff parametrization is limited in its applicability. At small $r$, the corresponding approximate dipole cross section
is, not surprisingly, inconsistent with the Soyez and GBW dipoles. However, the corresponding approximate dipole cross section for the
Block et al. parametrization agrees well with the Soyez dipole for $r$ less than about 1 GeV$^{-1}$ (for $\xi=x$). The approximate form of the
dipole cross section for small $r$, eq. (\ref{eq:emnlow}), also has a similar rise with increasing $r$.
We find that consistent with the Ewerz et al. discussion in Ref. \cite{ewerz}, unless the perturbative photon wave function is modified for low $Q^2$, the dipole cross
section should fall as $r$ becomes large,
contrary to what is assumed in many models.
While the low and high $r$ relationships between $F_2$ and dipole cross section discussed in Ref. \cite{ewerz} are only approximate, we find general concurrence between the approximate formulas and our $\hat{\sigma}(r,\xi)$.

The evaluation of $F_2(x,Q^2)$ in the dipole approach involves a convolution of the dipole cross section with the square of photon wave functions. The wave functions include the Bessel functions $K_0$ and $K_1$ which eventually fall exponentially. For large $Q^2$, the details of the large $r$ behavior of the dipole cross section are less important. Fig. 12 shows that $\hat{\sigma}$ approximated with eq. (\ref{eq:emnlow}), where $\xi=x$,
when integrated in eq. (5), is not far from the integral with the dipole cross section shown with the solid lines in Fig. 8.  

Our approach here, while only approximate, 
is complementary to much of the current literature where
a theoretical form of dipole cross section is postulated and its parameters are
fit to $F_2$ data. The functional form of the structure function itself is also postulated and fit to data.
Ultimately, one would like use methods similar to the ones discussed here to find dipole cross sections that represent a range of postulated $x$ and $Q$ dependencies in parametrization
of $F_2(x,Q^2)$ for further comparisons with theoretically motivated forms of the dipole cross section.

\appendix

\section{Donnachie-Landshoff \cite{dl} $F_2$}

The structure function $F_2(x,Q^2)$ parametrized by Donnachie-Landshoff is 
\be
F_2(x,Q^2)=Ax^{1-\alpha}\Biggl(\frac{Q^2}{Q^2+a}\Biggr)^\alpha
+ Bx^{1-\beta}\Biggl(\frac{Q^2}{Q^2+b}\Biggr)^\beta \,
\label{eq:f2dl}
\ee
where
\[ \begin{array}{c c c}
A=0.324 & a = 0.5616 \ {\rm GeV}^2 & \alpha=1.0808\\
B=0.098 & b=0.01114 \ {\rm GeV}^2\ & \beta = 0.5475\ .
\end{array}
\]
This is based on a fit to NMC data \cite{nmc} for $Q^2<10$ GeV$^2$. 
Although this parametrization is not valid for the full range, it is defined for the full range of $Q^2$, which is required for the evaluation of the Fourier transform.

\section{Block et al. \cite{blocknew} $F_2$}

The recent parametrization of $F_2$ by Block et al. \cite{blocknew} also has the property of being defined for the full range of $Q^2$. 
This Block el al. $F_2$ is valid for the wider $Q^2$ range and written as 

\begin{widetext}
\begin{eqnarray}
\label{eq:blocknewx}
F_2(x,Q^2) &=& D(Q^2)(1-x)^n\Biggl[ C(Q^2) + A(Q^2)
\ln \Biggl(\frac{1}{x}\frac{Q^2}{Q^2+\mu^2}\Biggr)
+B(Q^2)\ln^2 \Biggl(\frac{1}{x}\frac{Q^2}{Q^2+\mu^2}\Biggr)\Biggr]
\\
\label{eq:blocknewW}
\nonumber &=& D(Q^2)\Biggl(1-\frac{Q^2}{W^2+Q^2-m_p^2}\Biggr)^n\\ 
&\times & \Biggl[ C(Q^2) + A(Q^2) 
\ln \Biggl(\frac{W^2+Q^2-m_p^2}{Q^2+\mu^2}\Biggr)
+B(Q^2)\ln^2 \Biggl(\frac{W^2+Q^2-m_p^2}{Q^2+\mu^2}\Biggr)\Biggr] \ ,
\end{eqnarray}
where
\begin{eqnarray}
\nonumber A(Q^2) &=&a_0+a_1\ln\Biggl(1+\frac{Q^2}{\mu^2}\Biggr) + a_2  \ln^2
\Biggl(1+\frac{Q^2}{\mu^2}\Biggr)\\
B(Q^2) &=& b_0+b_1\ln\Biggl(1+\frac{Q^2}{\mu^2}\Biggr) + b_2  \ln^2
\Biggl(1+\frac{Q^2}{\mu^2}\Biggr)\\
\nonumber C(Q^2) &=& c_0+c_1\ln\Biggl(1+\frac{Q^2}{\mu^2}\Biggr)\ .
\end{eqnarray} 
\end{widetext} 
The parameters in these expressions are shown in Table 2. The function $D(Q^2)$ is
\begin{equation}
D(Q^2) = \frac{Q^2(Q^2+\lambda M^2)}{(Q^2+M^2)^2}\ .
\end{equation}
This fit to $F_2(x,Q^2)$ HERA data with the valence portion subtracted was done for 
$0.15$ GeV$^2\leq Q^2\leq 3000$ GeV$^2$, $W_b>25$ GeV and $x<0.1$. More details, including error bars on the fit parameter values, appears in Ref. \cite{blocknew}.

\begingroup
\squeezetable
\begin{table}[h]
\begin{center}
\vskip 0.25in
\begin{tabular}{|c|c|c|c|}
\hline
$ j$ &$ a_j$ &$ b_j$ & $c_j$  \\
\hline
\hline
0& $8.205\times 10^{-4}$ & 2.217$\times 10^{-3}$ & 0.255   \\
\hline
1& $-5.148\times 10^{-2}$ & 1.244$\times 10^{-2}$ & 1.475$\times 10^{-1}$    \\
\hline 
2& $-4.725\times 10^{-3}$ & $5.958\times 10^{-4}$ & $-$    \\
\hline
\hline
\end{tabular}
\caption{\label{table:paramb}
The central values of the parameter in $F_2$ parametrization of Block et al.\cite{blocknew}. Other parameters are $M^2=0.753$ GeV$^2$, $\mu^2=2.82$ GeV$^2$, $n=11.49$ and $\lambda = 2.430$.}\end{center}
\vspace{-0.6cm}
\end{table}
\endgroup

\section{Dipole cross sections}

The Golec-Biernat and W\"usthoff color dipole cross section is \cite{gbw}
\be
\hat{\sigma} _{GBW}= \sigma_0^{GBW} \Biggl[ 1-\exp\Biggl(
-\frac{r^2}{{4}r_c^2}\Biggl(\frac{x_0}{x}\Biggr)^\lambda\Biggr)\Biggr]
\ee
where $\sigma_0^{GBW}=23$ mb, $\lambda = 0.29$, $x_0=3\times 10^{-4}$ and $r_c=1$ GeV$^{-1}$.  This is often written in terms of
an $x$ dependent saturation scale $Q_s(x)=Q_0(x_0/x)^{\lambda/2}$ where $Q_0=1/r_c=1$ GeV.
The approximate functional form incorporates the observed geometric scaling behavior and
depends only on the one combined variable ${\cal T}_r= r Q_s$.
The limiting behavior at
small $r$ is therefore $\hat{\sigma} _{GBW}\simeq \sigma_0^{GBW} ({\cal T}_r/2)^2$, while at large $r$,
$\hat{\sigma} _{GBW}\to \sigma_0^{GBW}$.

In Ref. \cite{soyez}, Soyez writes the color dipole cross section
\begin{eqnarray}
\label{eq:sigmadip2}
\hat{\sigma} _{S} &=&  \sigma_0^{S}{\cal N}(rQ_s,Y  )\\
\nonumber
{\cal N}(r Q_s,Y)&=&
\begin{cases}
\N_0
\left(\dfrac{{\cal T}_r}{2}\right)^{2\gamma_\text{eff}(x,r)}, \qquad &\text{for }
{\cal T}_r<2 \\
1-\exp\left[-a \ln^2 (b{\cal T}_r)\right], & \text{for }
{\cal T}_r>2\ .
\end{cases}
\end{eqnarray}
in terms of the scaling variable ${\cal T}_r$, the saturation scale $Q_s(x)$ and where $Y=\ln(1/x)$. In this parametrization,
$\N_0=0.7$, $\gamma_s=0.738$, $\lambda=0.220$, $x_0=1.63\times 10^{-5}$ and $\sigma_0^{S}=
27.3$ mb.
The ``effective anomalous dimension'' in eq. (\ref{eq:sigmadip2}) is
\be
\gamma_\text{eff}(x,r)=\gamma_s+\frac{\ln (2/{\cal T}_r)}{\kappa \lambda Y}
\ee
with $\kappa=9.94$ \cite{bfkl}.  For $r=0.3$ GeV$^{-1}$ and $x=10^{-4}$, $\gamma_\text{eff}\simeq 0.84$, so $\hat{\sigma}
\sim r^{1.7}$ at small $r$ for this value of $x$.

\acknowledgments
We thank F. Halzen for useful conversations and O. Nachtmann for helful correspondence. MHR thanks the Aspen Center for Physics for their hospitality.
This work was supported in part by the Department of Energy contract DE-SC0010114.
The work of C. S. Kim and Y. S. Jeong was supported by the National Research Foundation of Korea (NRF) grant funded by the Korea government of the Ministry of Education, Science and Technology (MEST)
(No. 2011-0017430) and (No. 2011-0020333).


\begin{thebibliography}{99}
\bibitem{pdg}
  J.~Beringer {\it et al.}  [Particle Data Group Collaboration],
  Phys.\ Rev.\ D {\bf 86}, 010001 (2012).

\bibitem{pao}
  P.~Abreu {\it et al.}  [Pierre Auger Collaboration],
  Phys.\ Rev.\ Lett.\  {\bf 109}, 062002 (2012)
  [arXiv:1208.1520 [hep-ex]].

\bibitem{gaisser}
  T.~K.~Gaisser,
  Cambridge, UK: Univ. Pr. (1990) 279 p

\bibitem{lipari}
  See, for example,
  P.~Lipari,
  Astropart.\ Phys.\  {\bf 1}, 195 (1993).

\bibitem{gribov}
  V.~N.~Gribov and L.~N.~Lipatov,
  Sov.\ J.\ Nucl.\ Phys.\ {\bf 15}, 438 (1972).

\bibitem{altarelli}
  G.~Altarelli and G.~Parisi,
  Ncul.\ Phys.\ B {\bf 126}, 298 (1977).

\bibitem{dokshitzer}
  Y.~L.~Dokshitzer,
  Sov.\ Phys.\ JETP {\bf 46}, 641 (1977).

\bibitem{glr}
  L.~V.~Gribov, E.~M.~Levin, and M.~G.~Ryskin,
  Phys.\ Rep.\ {\bf 100}, 1 (1983).

\bibitem{mq}
  A.~H.~Mueller and J.~W.~Qiu,
  Nucl.\ Phys.\ B {\bf 268}, 427 (1986).

\bibitem{nikolaev}
  N.~N.~Nikolaev and B.~G.~Zakharov,
  Z.\ Phys.\ C {\bf 49} 607 (1991)

\bibitem{mueller}
  A.~H.~Mueller,
  Nucl.\ Phys.\ B {\bf 415}, 373 (1994)

\bibitem{gbw}
  K. Golec-Biernat and M. W\"usthoff,
  Phys.\ Rev.\ D {\bf 59}, 014017 (1998)
  [hep-ph/9807513].

\bibitem{sgbk}
  A.~M.~Stasto, K.~J.~Golec-Biernat and J.~Kwiecinski,
  Phys.\ Rev.\ Lett.\  {\bf 86}, 596 (2001)
  [hep-ph/0007192].

\bibitem{bgbk}
  J.~Bartels, K.~J.~Golec-Biernat, and H.~Kowalski,
	Phys.\ Rev.\ D {\bf 66}, 014001 (2002)
	[hep-ph/0203258].

\bibitem{iim}
  E.~Iancu, K.~Itakura, and S.~Munier,
	Phys.\ Lett.\ B {\bf 590}, 199 (2004)
	[hep-ph/0310338].

\bibitem{soyez}
  G. Soyez,
  Phys.\ Lett.\ B {\bf 655} (2007) 32-38.

\bibitem{albacete}
  J.~L.~Albacete \textit{et al.},
  Phys.\ Rev.\ D {\bf 80} (2009) 034031
  [arXiv:0902.1112].

\bibitem{aamqs}
  J.~L.~Albacete \textit{et al.},
  Eur.\ Phys.\ J.\ C {\bf 71} 1705 (2011)

\bibitem{forshaw}
  J.~R.~Forshaw, G.~Karley, and G.~Shaw,
  Phys.\ Rev.\ D. {\bf 60} (1999) 074012
  [hep-ph/9903341].

\bibitem{kowalski}
  H.~Kowalski and D.~Teaney,
  Phys.\ Rev.\ D. {\bf 68} (2003) 114005
  [hep-ph/:0304189].

\bibitem{gkmn}
  V.~P.~Goncalves \textit{et al.},
	Phys.\ Lett.\ B {\bf 643}, 273 (2006)
	[hep-ph/0608063].

\bibitem{dl}
  A. Donnachie and P.~V.~Landshoff
  Z.\ Phys.\ C {\bf 61}, 139-145 (1994).

\bibitem{haidt0}
  D.~Haidt
  Nucl.\ Phys.\ B, Proc.\ Suppl.\ {\bf 79}, (1999) 186-188.

\bibitem{block1}
  M.~M.~Block, E.~L.~Berger and C.~-I.~Tan,
  Phys.\ Rev.\ Lett. {\bf 97}, 252003 (2006).

\bibitem{block2}
  M.~M.~Block, E.~L.~Berger and C.~-I.~Tan,
  Phys.\ Rev.\ Lett. {\bf 98}, 242001 (2007).

\bibitem{block3}  
  M.~M.~Block, E.~L.~Berger, D.~W.~McKay and C.~-I.~Tan,
  Phys.\ Rev.\ D {\bf 77}, 053007 (2008).

\bibitem{block4}
  M.~M.~Block, P.~Ha and D.~W.~McKay,
  Phys.\ Rev.\ D {\bf 82}, 077302 (2010).

\bibitem{block5}
  M.~M.~Block, L.~Durand, P.~Ha and D.~W.~McKay,
  Phys.\ Rev.\ D {\bf 84}, 094010 (2011).

\bibitem{haidt}
  A.~Y.~Illarionov, B.~A.~Kniehl and A.~V.~Kotikov
  Phys.\ Rev.\ Lett. {\bf 106}, 231802 (2011).

\bibitem{block}  
  M.~M.~Block, L.~Durand, P.~Ha and D.~W.~McKay,
  Phys.\ Rev.\ D {\bf 88}, 014006 (2013)
  [arXiv:1302.6119].

\bibitem{blocknew}
 M.~M.~Block, L.~Durand and P.~Ha,
  arXiv:1404.4530 [hep-ph].

\bibitem{h197}
  C.~Adloff \textit{et al.}, [H1 collaboration],
  %
  Nucl.\ Phys.\ B {\bf 497} (1997) 3.

\bibitem{zeus99}
  J.~Breitweg \textit{et al.}, [ZEUS collaboration],
  Phys.\ Lett.\ B {\bf 407} (1997) 432;
  Eur.\ Phys.\ J.\ C {\bf 7} (1999) 609.

\bibitem{zeus00}
  J.~Breitweg \textit{et al.}, [ZEUS collaboration],
  Phys.\ Lett.\ B {\bf 487} (2000) 53
  [hep-ex/0005018].

\bibitem{h101}
  C.~Adloff \textit{et al.}, [H1 collaboration],
  Eur.\ Phys.\ J.\ C {\bf 21} (2001) 33
  [hep-ex/0012053].

\bibitem{zeus01}
  S.~Chekanov \textit{et al.}, [ZEUS collaboration],
  Eur.\ Phys.\ J.\ C {\bf 21} (2001) 443
  [hep-ex/0105090].

\bibitem{zeus}
  F.~D.~Aaron \textit{et al.}, [H1 and ZEUS collaboration],
  JHEP {\bf 01} (2010) 109
  [arXiv:0911.0884].

\bibitem{ewerz}
  C.~Ewerz, A.~von Manteuffel and O.~Nachtmann
  JHEP {\bf 03} (2011) 062
  [arXiv:1101.0288].

\bibitem{nmc}
  P.~Funaudrua \textit{et al.} [NMC collaboration],
	Phys.\ Lett.\ B {\bf 295} (1992) 159.

\bibitem{bfkl}
  L.~N.~Lipatov,
  Sov.\ J.\ Nucl.\ Phys.\ {\bf 23} 338 (1976);
	E.~A.~Kuraev, L.~N.~Lipatov, and V.~S.~Fadin,
	Sov.\ Phys.\ JETP {\bf 44} 443 (1976); {\bf 45} 199 (1977);
	I.~I.~Balitsky and L.~N.~Lipatov,
	Sov.\ J.\ Nucl.\ Phys.\ {\bf 28} 822 (1978).

\bibitem{nachtmann1}
  C.~Ewerz and O.~Nachtmann,
  Annals Phys.\  {\bf 322}, 1670 (2007)
  [hep-ph/0604087].

\bibitem{nachtmann2}
  C.~Ewerz and O.~Nachtmann,
  Phys.\ Lett.\ B {\bf 648}, 279 (2007)
  [hep-ph/0611076].


\bibitem{nachtmann3} 
  C.~Ewerz, A.~von Manteuffel and O.~Nachtmann,
  Phys.\ Rev.\ D {\bf 77}, 074022 (2008)
  [arXiv:0708.3455 [hep-ph]].





\end{thebibliography}
\end{document}